\documentclass[conference]{IEEEtran}
\usepackage{cite}
\usepackage{amsmath,amssymb,amsfonts}
\usepackage{algorithmic}
\usepackage{graphicx}
\usepackage{textcomp}
\usepackage[table,dvipsnames]{xcolor}      
\def\BibTeX{{\rm B\kern-.05em{\sc i\kern-.025em b}\kern-.08em
    T\kern-.1667em\lower.7ex\hbox{E}\kern-.125emX}}

\usepackage{booktabs}   
\usepackage{multirow}   
\usepackage{pbox}
\usepackage{adjustbox}
\usepackage{amsmath}
\usepackage{array}
\usepackage[frozencache,cachedir=./]{minted}
\usepackage{pifont}
\usepackage{hyperref}
\usepackage{array}
\usepackage{tabularx}
\usepackage{tikz}

\begin{document}
\newcommand{\bytes}{\mbox{bytes}}
\newcommand{\byte}{\mbox{byte}}
\newcommand{\second}{\mbox{s}}
\newcommand{\seconds}{\mbox{s}}
\newcommand{\flop}{\mbox{flop}}
\newcommand{\flops}{\mbox{flops}}
\newcommand{\NJFLOP}{\mbox{nJ/\flop}}
\newcommand{\instr}{\mbox{instr}}
\newcommand{\cycle}{\mbox{cy}}
\newcommand{\iter}{\mbox{it}}
\newcommand{\cycles}{\mbox{cy}}
\newcommand{\FCY}{\mbox{\flop/\cycle}}
\newcommand{\FIT}{\mbox{\flop/\iter}}
\newcommand{\BIT}{\mbox{\byte/\iter}}
\newcommand{\FR}{\mbox{\flops/\mbox{row}}}
\newcommand{\BR}{\mbox{\byte/\mbox{row}}}
\newcommand{\CYF}{\mbox{\cycles/\flop}}
\newcommand{\CS}{\mbox{\cycle/\second}}
\newcommand{\GCS}{\mbox{G\cycle/\second}}
\newcommand{\word}{\mbox{Word}}
\newcommand{\words}{\mbox{Words}}
\newcommand{\order}[1]{\mbox{${\cal O}\left(\mbox{#1}\right)$}}
\newcommand{\bit}{\mbox{bit}}
\newcommand{\bits}{\mbox{bits}}
\newcommand{\GBPS}{\mbox{G\bit/\second}}
\newcommand{\MBPS}{\mbox{M\bit/\second}}
\newcommand{\FS}{\mbox{\flop/\second}}
\newcommand{\BS}{\mbox{\byte/\second}}
\newcommand{\BC}{\mbox{\byte/\cycle}}
\newcommand{\GBS}{\mbox{G\byte/\second}}
\newcommand{\MBS}{\mbox{M\byte/\second}}
\newcommand{\GWS}{\mbox{G\word/\second}}
\newcommand{\GFS}{\mbox{G\flop/\second}}
\newcommand{\MFS}{\mbox{M\flop/\second}}
\newcommand{\lup}{\mbox{LUP}}
\newcommand{\lups}{\mbox{LUPs}}
\newcommand{\LUPCY}{\mbox{\lup/\cycle}}
\newcommand{\LUPS}{\mbox{\lup/\second}}
\newcommand{\MLUPS}{\mbox{M\lup/\second}}
\newcommand{\GLUPS}{\mbox{G\lup/\second}}
\newcommand{\GHZ}{\mbox{GHz}}
\newcommand{\ns}{\mbox{ns}}
\newcommand{\WF}{\mbox{\word/\flop}}
\newcommand{\BF}{\mbox{\byte/\flop}}
\newcommand{\FB}{\mbox{\flop/\byte}}
\newcommand{\BL}{\mbox{\byte/\lup}}
\newcommand{\GB}{\mbox{GB}}
\newcommand{\KB}{\mbox{kB}}
\newcommand{\MB}{\mbox{MB}}
\newcommand{\GiB}{\mbox{GiB}}
\newcommand{\MiB}{\mbox{MiB}}
\newcommand{\KiB}{\mbox{KiB}}
\newcommand{\TB}{\mbox{TB}}
\newcommand{\TiB}{\mbox{TiB}}
\newcommand{\W}{\mbox{W}}
\newcommand{\muarch}{\mbox{$\mu$-arch}}
\newcommand{\muop}{\mbox{$\mu$-op}}
\newcommand{\muops}{\mbox{$\mu$-ops}}

\newcommand{\eos}{\;.}
\newcommand{\cma}{\;,}
\newcommand{\rlm}{roof{}line model}
\newcommand{\rl}{roof{}line}
\newcommand{\Rlm}{Roof{}line model}
\newcommand{\Rl}{Roof{}line}
\newcommand{\olsep}{\|}
\newcommand{\nolsep}{|}
\newcommand{\ecmspace}{\,}
\newcommand{\TOL}{$T_{\mathrm{c}\_\mathrm{OL}}$}
\newcommand{\NNZR}{$N_\mathrm{nzr}$}
\newcommand{\NR}{$N_\mathrm{r}$}
\newcommand{\NNZ}{$N_\mathrm{nz}$}
\newcommand{\ecm}[6]{\mbox{$\left\{{#1}\ecmspace\olsep\ecmspace {#2}\ecmspace\nolsep\ecmspace {#3}\ecmspace\nolsep\ecmspace {#4}\ecmspace\nolsep\ecmspace {#5}\right\}\ecmspace{#6}$}}
\newcommand{\epsep}{\rceil}
\newcommand{\ecmp}[4]{\mbox{$\left\{{#1}\ecmspace\epsep\ecmspace {#2}\ecmspace\epsep\ecmspace {#3}\right\}\ecmspace{#4}$}}
\newcommand{\ecme}[4]{\mbox{$\left({#1}\ecmspace\epsep\ecmspace {#2}\ecmspace\epsep\ecmspace {#3}\right)\ecmspace{#4}$}}
\newcommand{\sellcs}{SELL-\texorpdfstring{$C$-$\sigma$}{C-sigma}}
\newcommand{\likwid}{\texttt{LIKWID}}
\newcommand{\likwidperfctr}{\texttt{likwid-perfctr}}
\newcommand{\likwidpin}{\texttt{likwid-pin}}
\newcommand{\likwidbench}{\texttt{likwid-bench}}
\newcommand{\lmbench}{\texttt{lmbench}}
\newcommand{\afx}{A64FX}
\newcommand{\spmv}{SpMV}
\newcommand{\cmg}{CMG}
\newcommand{\mve}{MVE}
\newcommand{\crs}{CRS}
\newcommand{\ellpack}{ELLPACK}
\newcommand{\tands}{dRECT}
\newcommand{\bq}{\begin{equation}}
\newcommand{\eq}{\end{equation}}
\newcommand{\cmark}{\textcolor{pastelgreen}{\ding{51}}}  
\newcommand{\xmark}{\textcolor{pastelred}{\ding{55}}}    
\newcommand{\ltype}[2]{\mbox{\texttt{(#1\,$\rightarrow$\,#2)}}}
\newcommand{\iform}[1]{\mbox{\texttt{#1}}}
\newcommand{\CC}{C\nolinebreak\hspace{-0.05em}%
  \raisebox{0.18ex}{\small\textbf{+}}\nolinebreak\hspace{-0.10em}%
  \raisebox{0.18ex}{\small\textbf{+}}\ }
\newcommand{\uarch}{$\mu$-arch}
\newcolumntype{P}[1]{>{\centering\arraybackslash}p{#1}}
\newcolumntype{N}{>{\centering\arraybackslash}m{1.5cm}}

\newcommand{\inputtikz}[2]{%
	\input{#1/#2.tex}%
}

\newenvironment{customlegend}[1][]{%
	\begingroup
	\csname pgfplots@init@cleared@structures\endcsname
	\pgfplotsset{#1}%
}{%
	\csname pgfplots@createlegend\endcsname
	\endgroup
}%
\def\addlegendimage{\csname pgfplots@addlegendimage\endcsname}

\makeatletter
\newcommand{\linebreakand}{%
  \end{@IEEEauthorhalign}
  \hfill\mbox{}\par
  \mbox{}\hfill\begin{@IEEEauthorhalign}
}
\makeatother

\newcommand{\GHacomm}[1]{{\color{red}-- {#1} --\color{black}} }
\newcommand{\JLcomm}[1]{{\color{orange}-- {#1} --\color{black}} }
\newcommand{\TRcomm}[1]{{\color{magenta}-- {#1} --\color{black}} }
\newcommand{\revon}{\color{black}}
\newcommand{\revoff}{\color{black}}


\definecolor{amber}{rgb}{1.0, 0.75, 0.0}
\definecolor{amethyst}{rgb}{0.6, 0.4, 0.8}
\definecolor{applegreen}{rgb}{0.55, 0.71, 0.0}
\definecolor{tumbleweed}{rgb}{0.87, 0.67, 0.53}
\definecolor{greyback}{RGB}{248,248,248}
\definecolor{deepblue}{RGB}{43,131,186}
\definecolor{greenalt}{RGB}{35,139,69}
\definecolor{darkred}{RGB}{215,25,28}
\definecolor{darkorange}{RGB}{253,174,97}
\definecolor{light-gray}{gray}{0.85}
\definecolor{lighter-gray}{gray}{0.95}
\definecolor{applegreen}{rgb}{0.55, 0.71, 0.0}
\definecolor{asparagus}{rgb}{0.53, 0.66, 0.42}
\definecolor{babyblueeyes}{rgb}{0.63, 0.79, 0.95}
\definecolor{burntsienna}{rgb}{0.91, 0.45, 0.32}
\definecolor{deepcarmine}{rgb}{0.66, 0.13, 0.24}
\definecolor{lightcornflowerblue}{rgb}{0.6, 0.81, 0.93}
\definecolor{steelblue}{rgb}{0.27, 0.51, 0.71}
\definecolor{pastelblue}{rgb}{0.68, 0.78, 0.81}
\definecolor{pastelbrown}{rgb}{0.51, 0.41, 0.33}
\definecolor{pastelgray}{rgb}{0.81, 0.81, 0.77}
\definecolor{pastelgreen}{rgb}{0.47, 0.87, 0.47}
\definecolor{pastelmagenta}{rgb}{0.96, 0.6, 0.76}
\definecolor{pastelorange}{rgb}{1.0, 0.7, 0.28}
\definecolor{pastelpink}{rgb}{1.0, 0.82, 0.86}
\definecolor{pastelpurple}{rgb}{0.7, 0.62, 0.71}
\definecolor{pastelred}{rgb}{1.0, 0.41, 0.38}
\definecolor{pastelviolet}{rgb}{0.8, 0.6, 0.79}
\definecolor{pastelyellow}{rgb}{0.99, 0.99, 0.59}
\definecolor{darkblue}{rgb}{0.0, 0.0, 0.55}

\definecolor{blue-seq-0}{RGB}{255, 255, 204}
\definecolor{blue-seq-1}{RGB}{119, 233, 180}
\definecolor{blue-seq-2}{RGB}{127, 205, 187}
\definecolor{blue-seq-3}{RGB}{ 65, 182, 196}
\definecolor{blue-seq-4}{RGB}{ 44, 127, 184}
\definecolor{blue-seq-5}{RGB}{ 37,  52, 148}
\definecolor{red-seq-0}{RGB}{254, 229, 217}
\definecolor{red-seq-1}{RGB}{252, 187, 161}
\definecolor{red-seq-2}{RGB}{252, 146, 114}
\definecolor{red-seq-3}{RGB}{251, 106, 74}
\definecolor{red-seq-4}{RGB}{222,  45, 38}
\definecolor{red-seq-5}{RGB}{165,  15, 21}
\definecolor{green-seq-0}{RGB}{237,248,233}
\definecolor{green-seq-1}{RGB}{199,233,192}
\definecolor{green-seq-2}{RGB}{161,217,155}
\definecolor{green-seq-3}{RGB}{116,196,118}
\definecolor{green-seq-4}{RGB}{49,163,84}
\definecolor{green-seq-5}{RGB}{0,109,44}
\definecolor{qual-0}{RGB}{102,194,165}
\definecolor{qual-1}{RGB}{252,141,98}
\definecolor{qual-2}{RGB}{141,160,203}
\definecolor{qual-3}{RGB}{231,138,195}
\definecolor{qual-4}{RGB}{166,216,84}
\definecolor{qual-5}{RGB}{255,217,47}
\definecolor{qual-6}{RGB}{229,196,148}
\definecolor{cbf-0}{RGB}{215,48,39}
\definecolor{cbf-1}{RGB}{244,109,67}
\definecolor{cbf-2}{RGB}{253,174,97}
\definecolor{cbf-3}{RGB}{254,224,144}
\definecolor{cbf-4}{RGB}{224,243,248}
\definecolor{cbf-5}{RGB}{171,217,233}
\definecolor{cbf-6}{RGB}{116,173,209}
\definecolor{cbf-7}{RGB}{69,117,180}

\definecolor{longqual-0}{RGB}{166,206,227}
\definecolor{longqual-1}{RGB}{31,120,180}
\definecolor{longqual-2}{RGB}{178,223,138}
\definecolor{longqual-3}{RGB}{51,160,44}
\definecolor{longqual-4}{RGB}{251,154,153}
\definecolor{longqual-5}{RGB}{227,26,28}
\definecolor{longqual-6}{RGB}{253,191,111}
\definecolor{longqual-7}{RGB}{255,127,0}
\definecolor{longqual-8}{RGB}{202,178,214}
\definecolor{longqual-9}{RGB}{106,61,154}
\definecolor{longqual-10}{RGB}{255,255,153}
\definecolor{longqual-11}{RGB}{177,89,40}

\title{WINIC: An Automatic Cycle-Accurate Multi-Platform Microbenchmarking Tool for Instruction Performance Metrics
\thanks{This work was partly funded by BMFTR through the DAREXA-F project.}
}

\author{
    \IEEEauthorblockN{Tobias Rühl}
    \IEEEauthorblockA{\textit{Erlangen National High Performance Computing Center} \\
    \textit{Friedrich-Alexander-Universität Erlangen-Nürnberg}\\
    Erlangen, Germany \\
    tobias.ruehl@fau.de}
\and
    \IEEEauthorblockN{Jan Laukemann}
    \IEEEauthorblockA{\textit{Erlangen National High Performance Computing Center} \\
    \textit{Friedrich-Alexander-Universität Erlangen-Nürnberg}\\
    Erlangen, Germany \\
    jan.laukemann@fau.de}
\linebreakand
    \IEEEauthorblockN{Georg Hager}
    \IEEEauthorblockA{\textit{Erlangen National High Performance Computing Center}\\
    \textit{Friedrich-Alexander-Universität Erlangen-Nürnberg}\\
    Erlangen, Germany \\
    georg.hager@fau.de}
\and
    \IEEEauthorblockN{Gerhard Wellein}
    \IEEEauthorblockA{\textit{Erlangen National High Performance Computing Center}\\
    Erlangen, Germany \\
    gerhard.wellein@fau.de}
}

\maketitle
\thispagestyle{plain}

\begin{abstract}
Performance models allow developers to gain important insights into the interaction of their code with the hardware that executes it, and enable them to apply target-specific optimizations.
To create in-core performance models for out-of-order processors, architecture-specific performance metrics like latency, throughput, and port binding of individual instructions are indispensable.
For x86 microarchitectures, there already exist tools to automatically obtain those values; however, the extension of performance modeling tools like the Open Source Architecture Code Analyzer (OSACA) to support further architectures like RISC-V emphasizes the need for a cross-platform microbenchmarking tool.
We present WINIC (What I Need Is Cycles), an automatic cross-platform microbenchmarking tool.
Unlike previous tools, it can obtain throughput and operand-based latency metrics for most supported instructions of any 64-bit x86, AArch64, and RISC-V microarchitecture without the need for root permissions or extended system control levels.
We cover the methodology of microbenchmarking and the approach of WINIC to collect cycle-accurate performance data for assembly instructions.
We show that WINIC can be used to automatically obtain a vast number of accurate results for x86, AArch64, and RISC-V microarchitectures. We show for four target microarchitectures that its results match between 71\% and 99\% of the performance data of instructions obtained by similar tools and documentation.
Furthermore, we show how operand-based metrics measured by WINIC can be used to improve the in-core performance predictions on selected micro-kernels, exemplarily for the Neoverse V2 microarchitecture.
\end{abstract}


\section{Introduction}\label{sec:introduction}

\subsection{Motivation}\label{ssec:motivation}
Optimizing large-scale scientific applications is critical for reducing their runtime and energy consumption. In this endeavor, performance modeling plays a vital role for understanding the interaction between the software and the hardware executing it. 
Although there exists a wide range of performance modeling tools, only so-called \emph{white-box} models~\cite{Eeckhout2010} reveal potential bottlenecks and allow users to gain insights on how to optimize their code. 
These tools require precise information on the underlying microarchitecture.
In-core execution modeling requires performance characteristics of machine instructions, which is often the reason for large performance databases within the software.
Such databases are tedious to create manually.
In the best case, they capture enough performance-related metrics to enable building a well-founded performance model.

In this work, we propose \emph{WINIC~(What I Need Is Cycles)}, a versatile microbenchmarking tool, designed to complement in-core performance modeling.
It  automatically measures throughput and operand-based latency on x86, AArch64, and RISC-V platforms while being able to run solely in user-space.
We gather data for four state-of-the-art microarchitectures, consisting of two x86 microarchitectures, namely the AMD Zen\,4 (``Genoa'') and Intel Ice Lake cores, the Arm-based Neoverse~V2 used in the Grace CPU Superchip, and the Spacemit~X60/K1 RISC-V microarchitecture used in the Banana~Pi~F3 CPU, and show that WINIC's measured results match 71--99\% of previously obtained performance numbers by other tools and references.
Furthermore, we apply the operand-based latency measurements to the in-core model of the Open Source Architecture Code Analyzer (OSACA)~\cite{OSACA1, OSACA2} and show how previously inaccurate predictions can be significantly improved by WINIC's input.

The rest of this paper is structured as follows: In Section~\ref{ssec:related_work}, we give an overview of the existing related work on this topic.
We discuss the design and performance metrics of modern out-of-order microarchitectures in Section~\ref{sec:background} and go into detail about the design and usage of WINIC in Section~\ref{sec:methodology}; in particular, the automated benchmarking of operand-based latencies.
Section~\ref{sec:results} evaluates the metrics obtained with WINIC by comparing them to existing data, and Section~\ref{sec:case_study} shows how these metrics can be used to improve OSACA's predictions.
Finally, in Section~\ref{sec:conclusion}, we summarize the paper and give an outlook on future work.

\subsection{Related Work}\label{ssec:related_work}
Although there is a variety of white-box basic-block predictors, such as the Intel Architecture Code Analyzer~(IACA)~\cite{IACA}, the LLVM Machine Code Analyzer~(llvm-mca)~\cite{llvm-mca}, the uops.info Code Analyzer~(uiCA)~\cite{uiCA}, \textsc{FACILE}~\cite{Facile}, the Code Quality Analyzer~(CQA)~\cite{cqa_rubial_2014}, or the Open Source Architecture Code Analyzer~(OSACA)~\cite{OSACA1, OSACA2}, they all require the underlying performance characteristics for executed instructions and can lack accuracy if this knowledge base is incomplete.
This is where microbenchmarking tools play an important role, as they provide the instruction-specific metrics on which these models are based.

nanoBench \cite{Abel_2020} is a microbenchmarking tool for x86-only.
It executes benchmarking kernels and then uses the \texttt{RDMSR} and \texttt{RDPMC} instructions to read core-local performance counters, which store the number of core cycles or of $\mu$ops dispatched on a specific port.
Those values are then used to determine throughput, latency, and port usage of the executed instructions.
A database with results for recent Intel and AMD microarchitectures is available at uops.info \cite{UopsPaper}.

iBench \cite{iBench} is a tool for measuring the latency and throughput of single instructions purely based on runtime and instruction count.
It can execute loop kernels manually written by the user and report the number of clock cycles elapsed per instruction. Currently, it only provides templates for x86 and AArch64.
WINIC is based on iBench, but adds support for RISC-V and generates the benchmark kernels automatically.

asmbench \cite{asmbench} is a benchmarking toolkit for x86 and AArch64.
It uses the LLVM just-in-time~(JIT) compilation via the llvmlite Python binding to generate and execute benchmarks.

llvm-exegesis \cite{exegesis} is a microbenchmarking tool for x86, AArch64, RISC-V, MIPS and PowerPC.
It can automatically generate and execute benchmarks for latency, throughput, and port usage; however, it cannot measure latencies per operand, which will be covered in Section~\ref{ssec:latency}.
Furthermore, it has no automatic mitigation of implicit dependencies in throughput benchmarks, covered in Section~\ref{ssec:meas_tp}.

The MAQAO \cite{maqao2005, maqao2019} framework that includes CQA also has a \emph{microbench} module to measure latency and throughput of x86 instructions. It is not publicly available so the exact capabilities at the time of writing are unclear.

\begin{table}[tp]
    \centering
    \caption{Capabilities of existing microbenchmarking tools compared to WINIC. $^{*}$ indicates only Intel support and $^\dagger$ the lack of documentation for this category.}
    \setlength\tabcolsep{5.05pt}
    \begin{adjustbox}{max width=\linewidth} 
    \begin{tabularx}{1.21\linewidth}{N|c|c|c|c|c|c}
        & \textbf{WINIC} & nanoBench & iBench & asmbench & \multicolumn{1}{N|}{llvm-exegesis} & MAQAO\\
        \hline
        x86        & \cmark & \cmark & \cmark & \cmark & \cmark & \cmark \\
        \rowcolor{lighter-gray}
        AArch64    & \cmark & \xmark & \cmark & \cmark & \cmark & \cmark~\scriptsize(beta) \\
        RISC-V     & \cmark & \xmark & \xmark & \xmark & \cmark & \xmark \\
        \rowcolor{lighter-gray}
        throughput & \cmark & \cmark & \cmark & \cmark & \cmark & \cmark \\
        latency    & \cmark & \cmark & \cmark & \cmark & \cmark & \cmark \\
        \rowcolor{lighter-gray}
        \multicolumn{1}{N|}{per-operand \newline latency} & \cmark & \cmark & \cmark & \xmark & \xmark & \xmark \\
        port usage & \xmark & \cmark\rlap{\hspace{0.2ex}$^{*}$} & \xmark & \xmark & \cmark\rlap{\hspace{0.2ex}$^{*}$} & \xmark \\
        \rowcolor{lighter-gray}
        automatic  & \cmark & \cmark & \xmark & \xmark & \cmark & \cmark \\
        user-space & \cmark & \cmark & \cmark & \cmark & \cmark & ?\rlap{\hspace{0.2ex}$^\dagger$} \\
        \rowcolor{lighter-gray}
        \multicolumn{1}{N|}{PMU\newline independent} & \cmark & \xmark & \cmark & \cmark & \xmark & \xmark \\
    \end{tabularx}
    \end{adjustbox}
    \label{tab:compare-tools}
\end{table}

Table~\ref{tab:compare-tools} compares the capabilities of state-of-the-art microbenchmarking tools.
Important features are the supported architectures (x86, AArch64, RISC-V) and metrics (latency, per-operand latency, throughput, port usage), as well as automatic benchmark generation, the possibility to run the tool in user-space mode, and the dependence on hardware performance counters.
While it is desirable to check all boxes in this comparison, no tool currently can provide this.
Although nanoBench can provide all capabilities for Intel architectures, the concept of using events of hardware performance units~(PMUs) for determining port bindings is not applicable to other CPU architectures and, therefore, limits the port derivation feature of this tool to Intel hardware.
WINIC aims for a different approach that can be transferred to all ISAs, using generic measurement techniques available to all platforms and we plan to extend its capabilities of port usage derivation soon.

\section{Background}\label{sec:background}

\subsection{Port Model}\label{ssec:port_model}

The out-of-order execution of modern processor cores can be simplified in a so-called \emph{port model}.
Each port represents one or multiple \emph{execution pipelines} responsible for a certain type of instruction, and each execution pipeline can run in parallel to other execution pipelines, while the scheduler can usually issue one instruction per port. 
Figure~\ref{fig:port_model} shows a simplified port model of a Neoverse~V2 core as an example of an out-of-order processor architecture. The instructions are decoded into one or multiple micro-operations ($\mu$ops), each of which is assigned to one of the 17~ports by the out-of-order scheduler.
As execution pipelines can exist in similar or the same form on multiple ports, the core can execute multiple instructions of the same kind in parallel.
The Neoverse~V2 core has two integer multiplication pipelines behind ports 6 and 7, respectively, allowing it to reach a peak throughput of 2\,instr/cy for this type of instruction.

Although some vendors publish such port models in their technical documentation, some vendors keep this information undisclosed, and, thus, microbenchmarking certain instructions in isolation as well as interleaved with other types of instructions can help uncover the anatomy of a core, which is crucial for creating an accurate performance model for it. 

\begin{figure}[tb!]\centering
    \includegraphics*[width=\linewidth]{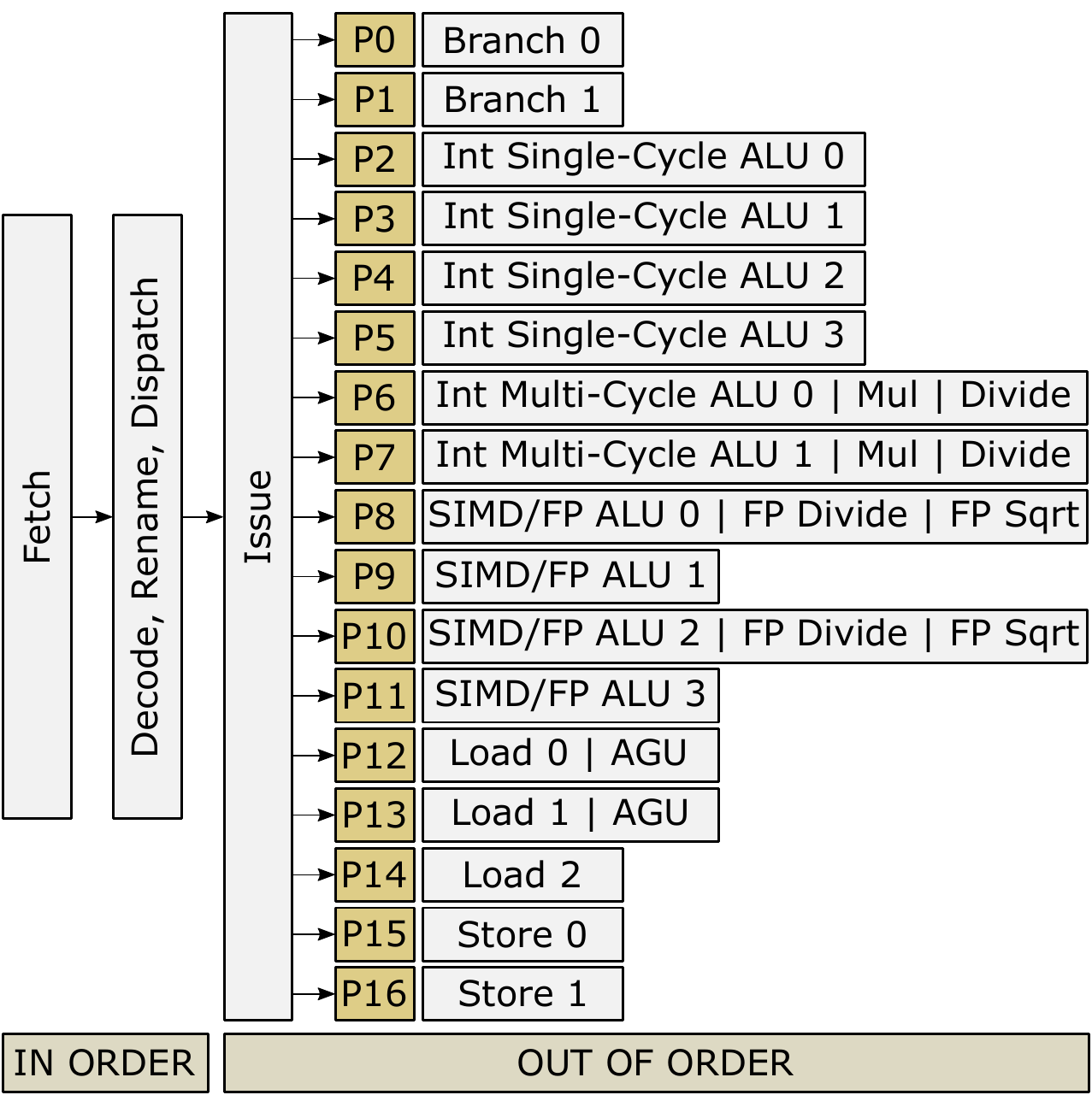}
    \caption{Arm Neoverse~V2 core block diagram and port model, compiled from the Arm Software Optimization Guide~\cite{NeoverseSOG}. The port numbering (``\texttt{P[0-9]+}'') is chosen by the authors and is not part of the official documentation.
    }
    \label{fig:port_model}
    \vspace{-10pt}
\end{figure}

\subsection{Latency}\label{ssec:latency}
For the following sections, it is vital to differentiate instructions with the same name but different operand types.
Therefore, aligning with previous related work~\cite{OSACA1}, we will use the term \textit{instruction form} to refer to an assembly instruction in combination with its operand types, while using the term \textit{instruction} when referencing an instance of an instruction form, e.g. in an assembly file, or a machine instruction that the CPU executes.
We will denote instruction forms as follows: \texttt{Mnemonic (Types...)}.
The minimum time from issue to retirement is referred to as the \emph{latency} of an instruction form.
When an instruction is issued, all operands that get written by the instruction are locked to ensure that no other instructions that may run concurrently can modify them.
Once the computation is finished, they are freed again and can be used by subsequent instructions.
However, on some instruction forms, some operands may be freed earlier than others.
For example, on x86 \iform{BTS\,(r16,\,r16)} uses the second operand as an index to select a bit in the first operand, stores it in the carry flag~(CF), and then sets it to $1$.
On AMD's Zen\,4 microarchitecture, this instruction form has a latency of 2~cycles, but the CF is already set and can be used by other instructions after $1$~cycle.
This is why the latency of the instruction form itself is not sufficient for an accurate performance prediction in some cases.
In the following, we will call the time between locking a read operand and freeing a written operand of an instruction form an \textit{operand-based latency} of the instruction form.
Section~\ref{ssec:measuring_latency} covers how WINIC measures those operand-based latencies.

\subsection{Throughput}
The number of independent instances of an instruction form that can be executed per clock cycle in a steady state is called the \emph{throughput} of that instruction form.
It can be limited by the number of ports that can execute it or the throughput of the sequential part of the pipeline, e.g., the retirement.
To maintain the same unit for throughput and latency, it is common to specify reciprocal throughput (cycles/instruction). In the following, when mentioning throughput values, reciprocal throughput is implied unless stated otherwise.

\subsection{The Open Source Architecture Code Analyzer~(OSACA)}\label{ssec:osaca}
OSACA is an in-core performance modeling tool.
This means that it assumes all memory accesses are L1 cache hits and focuses only on the in-core execution of a steady-state loop kernel.
Furthermore, it is assumed that the out-of-order scheduling works perfectly and the front end does not depict any bottleneck.
Under those assumptions, there are two main factors that can limit the performance of a loop kernel: i) The throughput of the port with the highest \textit{port pressure}, i.e., the port with the longest accumulated execution time per loop iteration, or ii) data dependencies between instructions.
As long as there are only data dependencies between instructions of one loop iteration, the latency of the \emph{critical path}, i.e., the longest inter-loop dependency chain, can usually overlap with critical paths of other iterations if the reorder buffer is large enough to hold a sufficient amount of unrolled loop iterations.
Dependencies across different loop iterations~(\textit{loop-carried dependencies}), however, impose another lower bound on the overall runtime.
One can predict the runtime derived from OSACA's analysis by taking the maximum out of the mentioned factors.
For a thorough understanding of the performance model and potential optimizations, OSACA provides the throughput of all individual ports, certain pipelines~(e.g., the divider pipeline), all loop-carried dependencies, and the critical path in its analysis output.

\begin{figure*}[tp]
    \centering
    \begin{adjustbox}{max width=\linewidth}
    \begin{tabular}{|r|*{19}{c|}l|}
    \hline
    line  & 0 & 1 & 2 & 3 & 4 & 5 & 6 & 7 & 8 & 9 & 10 & 11 & 12 & 13 & 14 & 15 & 16 & CP & LCD & Instruction \\
    \hline
    37 &  &  &  &  &  &  &  &  &  &  &  &  &  &  &  &  &  &  &  & \texttt{.LBB0\_9:} \\
    38 &  &  &  &  &  &  &  &  & 0.25 & 0.25 & 0.25 & 0.25 &  &  &  &  &  &  &  & \texttt{mov     v4.16b, v1.16b} \\
    39 &  &  & 0.25 & 0.25 &  &  & 0.25 & 0.25 &  &  &  &  & 0.33 & 0.33 & 0.33 &  &  & 4.0 &  & \texttt{ldr     q2, [x10], \#16} \\
    40 &  &  & 0.25 & 0.25 &  &  & 0.25 & 0.25 &  &  &  &  & 0.17 & 0.16 & 0.66 &  &  &  &  & \texttt{ldr     q3, [x11], \#16} \\
    41 &  &  & 0.33 &  &  &  & 0.33 & 0.33 &  &  &  &  &  &  &  &  &  &  &  & \texttt{subs    x13, x13, \#2} \\
    42 &  &  &  &  &  &  &  &  & 0.25 & 0.25 & 0.25 & 0.25 &  &  &  &  &  & 4.0 &  & \texttt{fmla    v4.2d, v3.2d, v2.2d} \\
    43 &  &  & 0.17 & 0.49 &  &  & 0.17 & 0.17 &  &  &  &  & 0.50 & 0.50 &  & 0.50 & 0.50 & 0.0 & 1.0 & \texttt{str     q4, [x12], \#16} \\
    44 & 0.50 & 0.50  &  &  &  &  &  &  &  &  &  &  &  &  &  &  &  &  &  & \texttt{b.ne    .LBB0\_9} \\
    \hline
      sum & 0.50 & 0.50 & 1.00 & 0.99 &  &  & 1.00 & 1.00 & 0.50 & 0.50 & 0.50 & 0.50 & 1.00 & 1.00 & 1.00 & 0.50 & 0.50 & 8.0 & 1.0 & \\
    \hline
    \end{tabular}
    \end{adjustbox}
    \caption{OSACA output example: for each instruction it shows the port pressure, its contribution to the critical path, and loop-carried dependencies if present. All throughput values are reciprocal throughput, i.e., lower is better.}
    \label{fig:osaca_x86}
\end{figure*}

Figure~\ref{fig:osaca_x86} shows an example of an OSACA analysis on the STREAM Triad~\cite{McCalpin1995} kernel as depicted in Listing~\ref{lst:STREAM}:

\begin{listing}[h]
\vspace{-1em}
\begin{minted}
[
xleftmargin=20pt,
frame=lines,
framesep=2mm,
baselinestretch=1.2,
bgcolor=lighter-gray,
fontsize=\footnotesize,
linenos
]
{cpp}
double a[N], b[N], c[N], d;

for(int i=0; i<N; ++i) {
    a[i] =  b[i] + c[i] * d;
}
\end{minted}
\vspace{-2em}
\caption{The STREAM triad.}
\label{lst:STREAM}
\end{listing}

It was compiled using Clang 21.1.0 with flags \texttt{-O3 -fno-unroll-loops}, and analyzed
for the Neoverse~V2 architecture.
Looking at the output, the ports with the highest port pressure~(indicated by the sum in the last line) are ports 2, 3, 6, 7, 12, 13, and 14, each with a throughput of $1$~cycle, and the loop-carried dependency (column ``LCD'') is $1$~cycle.
Therefore, the overall performance is bound by the saturation of such ports, which, according to Figure~\ref{fig:port_model}, are the ports having an ALU~(ports 2, 3, 6, 7) and a LOAD unit~(ports 12, 13, 14).
The critical path (column ``CP'') is reported to be $8$~cycles.
In summary, this output shows how the code interacts with the hardware of this specific architecture, and helps to identify the bottleneck.
\section{Methodology}\label{sec:methodology}
\subsection{Platform-Independent Approach}\label{ssec:approach}
To obtain reproducible performance metrics of a single instruction, it has to be executed repeatedly until a steady state is reached and external overhead by the loop code is overcome.
In addition, an isolated benchmarking environment is important to minimize interference of other processes.
There are several options to obtain reliable performance metrics from these measurements.
nanoBench \cite{Abel_2020} queries hardware performance counters of the CPU which report the number of instructions retired since the start of the measurement, as well as the instructions retired per port.
However, the counters for port occupation are only available on Intel CPUs, and while AArch64 and RISC-V provide performance counters to obtain the total number of instructions retired, to reduce platform-specific code as much as possible, we decided to rely purely on the total runtime of the loop instead.
The number of cycles per instruction can be calculated using the total runtime, number of instructions executed, and the clock frequency.
This is the same approach used by iBench \cite{iBench}.
WINIC can be seen as an extension of iBench, generating the loop body for throughput and latency benchmarks for every instruction form automatically.

\subsection{Measuring Throughput}\label{ssec:meas_tp}
To measure the throughput of an instruction form, a loop body with a sequence of those instructions has to be generated in a way that allows the CPU to parallelize them as much as possible.
Therefore, dependencies between individual instructions must be avoided.
The trivial way to do this is to choose different registers for each operand of each instruction. However, since the number of registers available is limited, this sometimes leads to very short sequences, which can have a negative impact on the accuracy of the measurements as it increases the loop overhead.
To generate more independent instructions, WINIC reuses registers for operands that are read but not written by the instruction form.
For example, \iform{VADDPD\,(r128,\,r128,\,r128)} on x86 writes to the first register operand and reads from the second and third operands.
WINIC generates the loop body shown in Listing \ref{lst:vaddpd}, reusing registers where possible.

\begin{listing}[h]
\begin{minted}
[
xleftmargin=20pt,
frame=lines,
framesep=2mm,
baselinestretch=1.2,
bgcolor=lighter-gray,
fontsize=\footnotesize,
linenos
]
{asm}
loop_tp:
  inc    i
  vaddpd  xmm0, xmm1, xmm2
  vaddpd  xmm3, xmm1, xmm2
  vaddpd  xmm4, xmm1, xmm2
  ...
  vaddpd  xmm12, xmm1, xmm2
  vaddpd  xmm13, xmm1, xmm2
  cmp    i, N
  jl     loop_tp
\end{minted}
\vspace{-2em}
\caption{The benchmark kernel generated by WINIC to measure the throughput of \iform{VADDPD\,(r128,\,r128,\,r128)}.}
\label{lst:vaddpd}
\end{listing}

Note that disjoint registers were chosen for the second and third operand within a single instruction.
This is because instruction forms can have semantics that allow for optimizations if two operands are the same, e.g. an \iform{VPXOR\,(r128, r128, r128)} will always have the result $0$ if the second and third operand are identical.
Modern processors can detect such cases and skip part of the pipeline to speed up execution, which would render the measurements useless.
The target number of instructions to generate can be adjusted with a CLI option with a default of 12 as this proves to yield reliable results while keeping the loop body of manageable size for manual adjustments.
In case there are not enough registers available, WINIC instead generates as many instructions as possible.

\subsubsection{Implicit Dependencies}\label{sssec:impl_deps}
Since instruction forms can have side effects, such as writes to a flag register, some of them have implicit dependencies on themselves regardless of the explicit registers chosen.
In order to measure the throughput of such instruction forms, we interleave other instructions to break those dependencies.
The interleaved instructions need to write to the register causing the dependency without introducing any other explicit or implicit dependencies into the benchmark kernel.
Based on those constraints, WINIC dynamically chooses the instruction form to break the dependency.
The \texttt{ADC} instruction forms on x86 are an example of this, as they read and write the carry-flag.
Even if choosing different registers for all instructions in the loop, they would still have an implicit dependency on each other.
WINIC detects this and automatically inserts a \texttt{TEST} between each of the \texttt{ADC}s as shown in Listing \ref{lst:interleave}.

\begin{listing}[h]
\begin{minted}
[
xleftmargin=20pt,
frame=lines,
framesep=2mm,
baselinestretch=1.2,
bgcolor=lighter-gray,
fontsize=\footnotesize,
linenos
]
{asm}
loop_tp:
  inc    i
  adc    rax, 7
  test   rcx, rdx
  adc    rsi, 7
  test   rcx, rdx
  ...
  adc    r13, 7
  test   rcx, rdx
  cmp    i, N
  jl     loop_tp
\end{minted}
\vspace{-2em}
\caption{Benchmark kernel generated to measure the throughput of \iform{ADC\,(r64,\,i32)}.}
\label{lst:interleave}
\end{listing}

Each \texttt{TEST} writes to the flags but does not depend on any \texttt{ADC} or \texttt{TEST} preceding it, assuming the registers are chosen correctly.
Therefore, it breaks the dependency and allows multiple \texttt{ADC} instructions to be executed in parallel.
However, adding those breaking instructions can affect the measurements.
If the interleaved instruction form uses the same execution units as the target instruction form, its throughput adds up to the overall runtime.
Therefore, WINIC reports a range within which the throughput must lie.
In case of the \texttt{ADC} benchmark, the measured throughput of a set of instructions $\mathit{TP}_\textrm{m}$ on Zen\,4 is $0.5$\,cycles.
In the best case, both instruction forms use the exact same ports, i.e., have to be executed without any overlap, so the throughput of the \texttt{ADC} instruction form can be calculated as 
$\mathit{TP}_\textrm{ADC} = \mathit{TP}_\textrm{m} - \mathit{TP}_\textrm{TEST}$.
With \texttt{TEST} having a throughput of $0.25$\,cycles, this would lead to $\mathit{TP}_\textrm{ADC} = 0.25\,\textrm{cy}$.
In the worst case, the execution of \texttt{ADC} can fully overlap with \texttt{TEST} and, therefore, takes up the full $0.5$\,cycles of the measurement.
We can infer that the throughput of \texttt{ADC} lies between $0.25$ and $0.5$\,cycles.

\subsection{Measuring Latency}\label{ssec:measuring_latency}
As discussed in Section~\ref{ssec:latency}, we have to measure all operand-based latencies of an instruction form individually.
For this, we classify the them according to the types of the source operand and the destination operand.
We call the combination of those the \textit{type} of the operand-based latency and denote it as \ltype{type(source)}{type(destination)}.

\subsubsection{Measuring Symmetric Latency}
We will call an operand-based latency type \textit{symmetric} if the source operand and the destination operand have the same type.
The \iform{VADDPD\,(r128,\,r128,\,r128)} instruction form has two symmetric operand-based latencies of type \ltype{r128}{r128} as it writes to the first operand and reads from both the second and third operands.
For the first operand-based latency, WINIC generates an assembly loop such as the one shown in Listing~\ref{lst:lat_vaddpd}.

\begin{listing}[]
\begin{minted}
[
xleftmargin=20pt,
frame=lines,
framesep=2mm,
baselinestretch=1.2,
bgcolor=lighter-gray,
fontsize=\footnotesize,
linenos
]
{asm}
loop_lat:
  inc     i
  vaddpd  xmm0, xmm1, xmm2
  vaddpd  xmm3, xmm0, xmm2
  vaddpd  xmm4, xmm3, xmm2
  ...
  vaddpd  xmm12, xmm11, xmm2
  vaddpd  xmm1, xmm12, xmm2
  cmp     i, N
  jl      loop_lat
\end{minted}
\vspace{-2em}
\caption{The benchmark kernel generated by WINIC to measure the operand-based latency of \iform{VADDPD\,(r128,\,r128,\,r128)} from the second operand to the first operand.}
\label{lst:lat_vaddpd}
\end{listing}

It chooses the registers for the two operands involved in a way that introduces a read-after-write dependency between the instructions, which limits the throughput of the kernel to the operand-based latency of this read-write operand pair.
The other registers are chosen in such way that no additional dependencies are introduced.

\subsubsection{Measuring Asymmetric Latency}\label{sssec:asym_lat}
To measure asymmetric operand-based latencies a different approach is needed.
For example, an operand-based latency of type \ltype{r64}{flags} cannot be measured alone; however, one can measure a combination of two instruction forms with operand-based latencies of type \ltype{r64}{flags} and \ltype{flags}{r64} by interleaving them and thus creating a latency chain.
This leaves one with some latency $L_\textrm{combined}$ and no information on which instruction form contributed how much to the combined result.
To narrow down the ranges of the individual latencies, we first group all asymmetric operand-based latency measurements by their types.
For a pair of complementary types \ltype{a}{b} and \ltype{b}{a}, all combinations of instructions from those types can be measured.
We then systematically measure those combinations until we find the one with the minimal combined latency $L_\textrm{combined, min}$.
The two instruction forms determined this way are then used to measure all other instruction forms with operand-based latencies of the two types.
For type pairs where $L_\textrm{combined, min}$ is $2$\,cycles, it can be inferred that both instruction forms have an operand-based latency of $1$\,cycle, as we assume every instruction needs at least $1$\,cycle to be executed, and the other operand-based latencies of the two types can be determined exactly.
This is the case for around 71\% of all latency values on Zen\,4, with 29\%  remaining ranges.
The vast majority of the instruction forms with latency ranges are AVX512 instructions including mask registers, as a dependency chain between the mask registers~(\texttt{k}) and vector registers~(\texttt{xmm}, \texttt{ymm}, \texttt{zmm}) can only be built with a limited set of instructions.
By assuming the respective vector-to-mask and mask-to-vector instructions both take an equivalent time, the number of instructions with ranged latencies could be significantly reduced.

An example of an instruction form with asymmetric operand-based latencies is \iform{BTS\,(r16,\,r16)} on Zen\,4.
It reads and writes the first register operand, reads the second register operand and writes the CF flag, therefore, it has four operand-based latencies, two of type \ltype{r16}{r16} and two of type \ltype{r16}{flags}.
To measure the operand-based latencies of type \ltype{r16}{flags}, another instruction form is needed.
On Zen\,4, WINIC finds the pair \iform{ADC\,(r16,\,r16)} and \iform{ADC\,(r16,\,i16)} with operand-based latencies of types \ltype{r16}{flags} and \ltype{flags}{r16} that have a combined latency of $2$\,cycles.
Thus, WINIC infers that both have a latency of $1$\,cycle, and uses them as helpers for other instruction forms, such as \iform{BTS\,(r16,\,r16)}.
It then generates the following loop to measure the operand-based latency of \iform{BTS\,(r16,\,r16)} from the second register operand to the flags:

\begin{listing}[H]
\begin{minted}
[
xleftmargin=20pt,
frame=lines,
framesep=2mm,
baselinestretch=1.2,
bgcolor=lighter-gray,
fontsize=\footnotesize,
linenos
]
{asm}
loop_lat:
  inc  i
  bts  ax, cx  # read cx, write flags
  adc  dx, si  # read flags, write dx
  bts  bx, dx  # read dx, write flags
  adc  r9w, si  # read flags, write r9
  ...
  bts  r12w, r15w  # read r15, write flags
  adc  cx, si  # read flags, write cx
  cmp  i, N
  jl   loop_lat

\end{minted}
\vspace*{-0.5cm}
\caption{The benchmark kernel generated for \iform{BTS\,(r16,\,r16)} to measure the operand-based latency from the second register operand to the flags.}
\label{lst:lat_bts}
\end{listing}
\vspace*{0.2cm}
There is a read-after-write dependency introduced by the register operands between each \texttt{ADC} and the next \texttt{BTS}.
Between each \texttt{BTS} and the next \texttt{ADC}, there is an implicit read-after-write dependency, as \texttt{BTS} writes to the flags and \texttt{ADC} reads them.
The latency of this operand-dependency is the one we are interested in and can be derived from the overall runtime and the known value of the injected operand-based latency.
The measurements reveal that, on Zen\,4, it is only $1$\,cycle, whereas the operand-based latencies of type \ltype{r16}{r16} are $2$\,cycles.

\subsection{Generating Instructions}
To automatically generate the benchmarking kernels with all correction mechanisms described above, a great extent of meta-information is needed, such as (i) a list of all supported instruction forms, (ii) an operand list and side effects per instruction form, (iii) the type and applicable values for each operand, and (iv) information on how to generate syntactically correct assembly code from the above.

LLVM \cite{LLVM} was chosen as an underlying toolsuite for this, as it can serve as a single source for all information needed for our use case.
LLVM is a "collection of modular and reusable compiler and toolchain technologies" \cite{LLVM}, including the Clang compiler, the LLDB debugger, implementations of the C and \CC standard libraries, and many more.
It supports a wide variety of architectures, making it ideal for the purpose of a multi-platform tool.
With the exception of memory operands, LLVM's internal data structures for storing this information are the same for all architectures, so it is possible to implement the loop generation logic in a mostly architecture-independent way.
Furthermore, LLVM is open source, well maintained, and is continuously updated to support new microarchitectures.

To generate the kernel in Listing~\ref{lst:interleave}, WINIC interacts with LLVM in the following way:
\begin{itemize}
    \item It looks up the operand list.
    \item For each register operand, it looks up all registers of the correct type for that operand.
    \item It generates the desired number of instructions, using LLVM's \texttt{MCInst} data structure and populates register and immediate operands, keeping track of the registers used to avoid dependencies.
    \item It detects the implicit dependency on the flags based on the list of implicit register usages LLVM provides for the instruction form and searches for a dependency-breaking instruction form.
    \item It generates suitable instructions for breaking dependencies and interleaves them.
    \item It uses LLVM's \texttt{MCInstrPrinter} to generate valid assembly from the set of \texttt{MCInst} instructions.
\end{itemize}

\subsection{Executing Benchmarks}
The generated kernels are embedded in a target-specific template that provides a function label, the loop logic, code to save and restore callee-saved registers, and code for initializing registers with specific values.

The result is assembled to a shared library, which is then loaded and provides the generated function with the loop count as its only argument.
Finally, the function is executed and its runtime measured.
Every benchmark runs in a subprocess to recover from faults or unwanted side effects that executing arbitrary instructions can result in.
To ensure reproducibility when assembling the benchmarks, a fixed version of Clang is used, which is built alongside LLVM during the setup process.

\subsection{Improving Accuracy}\label{ssec:improving_accuracy}
The total execution time of the benchmark function contains overhead from the loop mechanics, as well as the saving and restoring of registers.
To improve the accuracy of the measurements, WINIC does a second benchmark with the loop body unrolled twice.
Using the two results~($T_\mathrm{1}$ and $T_\mathrm{2xUNRL}$), the overhead can be arithmetically removed, with $T_{\mathrm{body}}$ describing the execution time needed for the loop body and $T_{\mathrm{LM}}$ representing the time for the loop mechanics:

\begin{equation}\label{eq:overhead}
\begin{split}
    T_\mathrm{2xUNRL} - T_\mathrm{1} 
    & = ( 2 * T_\mathrm{body} + T_\mathrm{LM}) - (T_\mathrm{body} + T_\mathrm{LM}) \\
    & = T_\mathrm{body}
\end{split}
\end{equation}

Some instructions can be executed efficiently if the operands meet certain criteria, e.g., the CPU may diverge from the normal execution path for a multiplication if one of the operands is zero.
To avoid such optimizations, all registers by default are initialized to uncritical values (neither $0$ or $1$).

On some x86 architectures, there is a performance penalty when transitioning between AVX and SSE instructions \cite{IntelAVX}.
The penalty occurs if a register previously used by an SSE instruction is used by an AVX instruction, and vice versa.
Since it is not guaranteed to occur both when running the normal and the unrolled benchmark, it cannot be corrected using Equation~\ref{eq:overhead} reliably.
To prevent a potential transition at the first loop iteration of the benchmark, an additional warmup run is done before the actual benchmark, setting the AVX/SSE state of all vector registers used.

\section{Results}\label{sec:results}
\subsection{Testing Environment}\label{ssec:test_env}
The results covered in this chapter were obtained on four systems, as listed in Table~\ref{tab:systems}, which allows us to verify WINIC's multi-platform capabilities on two x86~(Intel and AMD), one AArch64~(NVIDIA), and one RISC-V~(Banana Pi) models.

\begin{table*}[t]
    \centering
    \caption{Hardware and software environment for the conducted benchmarks in this paper.}
    \begin{adjustbox}{max width=\linewidth}
    \begin{tabular}{cl||cc|c|c}
          && \multicolumn{2}{c|}{x86-64} & AArch64 & RISC-V \\
    \hline
    \parbox[t]{2mm}{\multirow{4}{*}{\rotatebox[origin=c]{90}{HW}}}
    & CPU Model         & AMD EPYC 9654 & Intel Xeon Platinum 8360Y & NVIDIA Grace & Banana Pi F3 \\
    & Microarchitecture & Zen\,4  & Sunny Cove & Neoverse V2 & Spacemit X60/K1 \\
    & Base Frequency   & 2.4\,\GHZ & 2.4\,\GHZ & 3.1\,\GHZ & 1.6\,\GHZ \\
    & Fixed Frequency   & 1.5\,\GHZ & 2.4\,\GHZ & 3.2\,\GHZ & 1.6\,\GHZ \\
    \hline
    \parbox[t]{2mm}{\multirow{3}{*}{\rotatebox[origin=c]{90}{SW}}}
    & Compiler & GCC 13.3.0 & GCC 13.3.0 & GCC 13.3.0 & GCC 15.1.1 \\
    & OS & Ubuntu 24.04 & Ubuntu 24.04 & Ubuntu 24.04 & ArchLinux (rolling) \\
    & Linux kernel & 6.12.115 & 6.12.115 & 6.8.0 & 6.1.15 \\
    \end{tabular}
    \end{adjustbox}
    \vspace{2pt}
    \label{tab:systems}
\end{table*}

WINIC\footnote{Available at \url{https://github.com/RRZE-HPC/WINIC}} uses LLVM and Clang in version 22.1.8 on all platforms.
At the time of writing, there was no single version of the GCC compiler available on all four systems, which is why we used a different version on the RISC-V system.
The measurements for this chapter were done using the WINIC repository in commit \href{https://github.com/RRZE-HPC/WINIC/tree/ae34a899347fb5c27eb651e5ad82c1a2424ae227}{\texttt{ae34a89}}.
We use llvm-exegesis from LLVM 22.1.8 with flags \texttt{--repetition-mode=loop --loop-body-size=12 --min-instructions=12000000} and OSACA in its most current release of version 0.7.1.
The measurements provided by uops.info date back to the most current database version at the time of the writing~(March 2026).

\subsection{Operand-based Latency Statistics}\label{ssec:statistics}
In section \ref{ssec:latency} we showed that for each combination of read and written operands an instruction form can potentially have a different latency.
Figure \ref{fig:statistics} shows what proportion of the instruction forms measured on our test systems actually show this behavior.
An instruction form can either have 1) only one read/write operand combination to begin with (shown in red), 2) multiple combinations measured to have the same operand-based latency (marked in orange) or 3) multiple distinct operand-based latencies (shown in blue).
Marked in gray are cases where WINIC reports a latency range for at least one operand-based latency that overlaps with the other ones, and, therefore, it is not clear if the precise values would be the same.
For the x86 systems, around 40\% of instruction forms have different operand-based latencies in our measurements.
On the Arm and RISC-V systems, this proportion is only 14\% and 10\%, respectively.

\begin{figure}
    \centering
    \includegraphics[width=\linewidth]{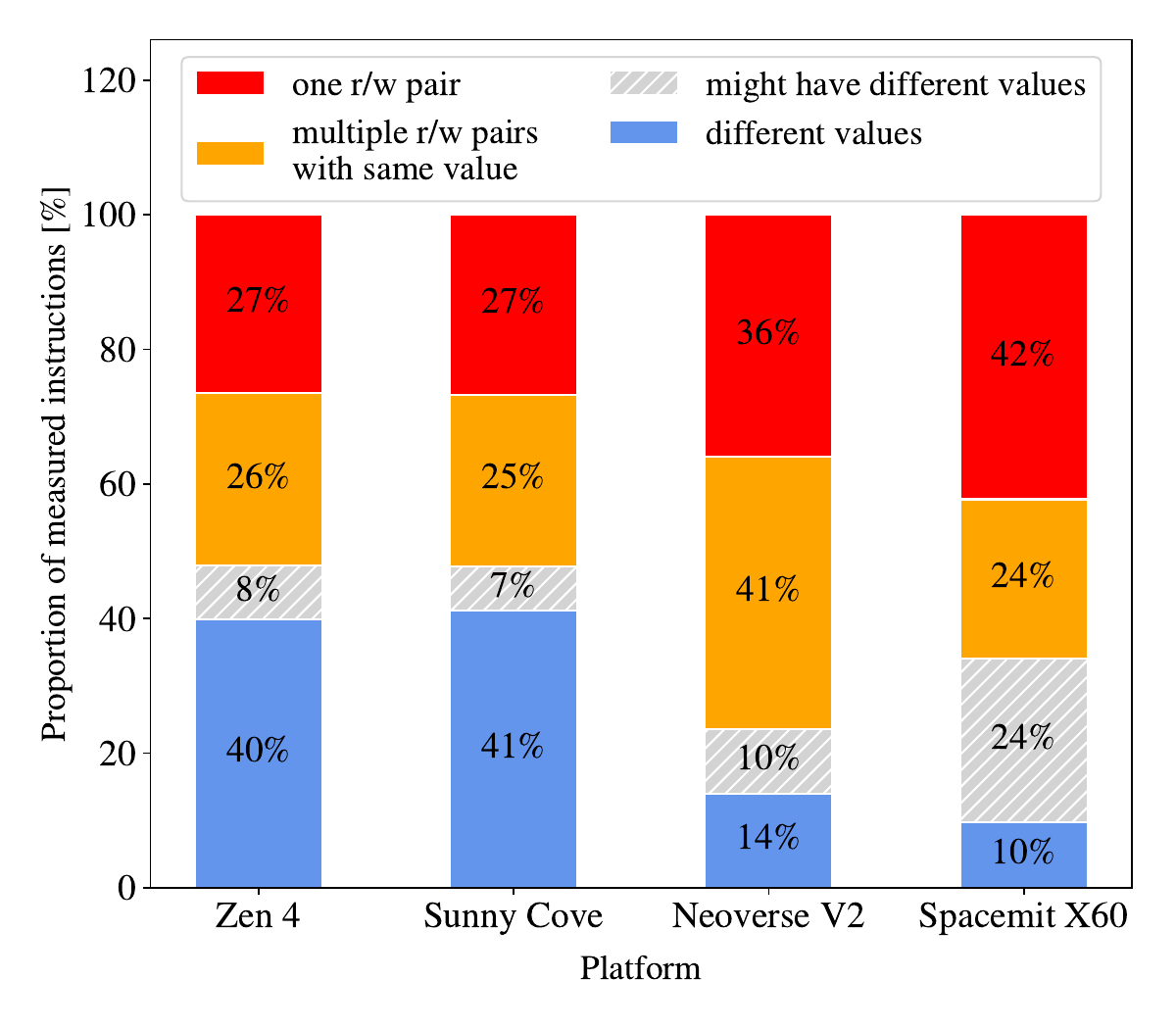}
    \vspace*{-0.9cm}
    \caption{Classification of instructions with operand-based latencies on the investigated microarchitectures.}
    \label{fig:statistics}
\end{figure}

\subsection{Comparison to Existing Data}\label{ssec:comparison}
\begin{figure*}[tp]
    \centering
    \includegraphics[width=\linewidth]{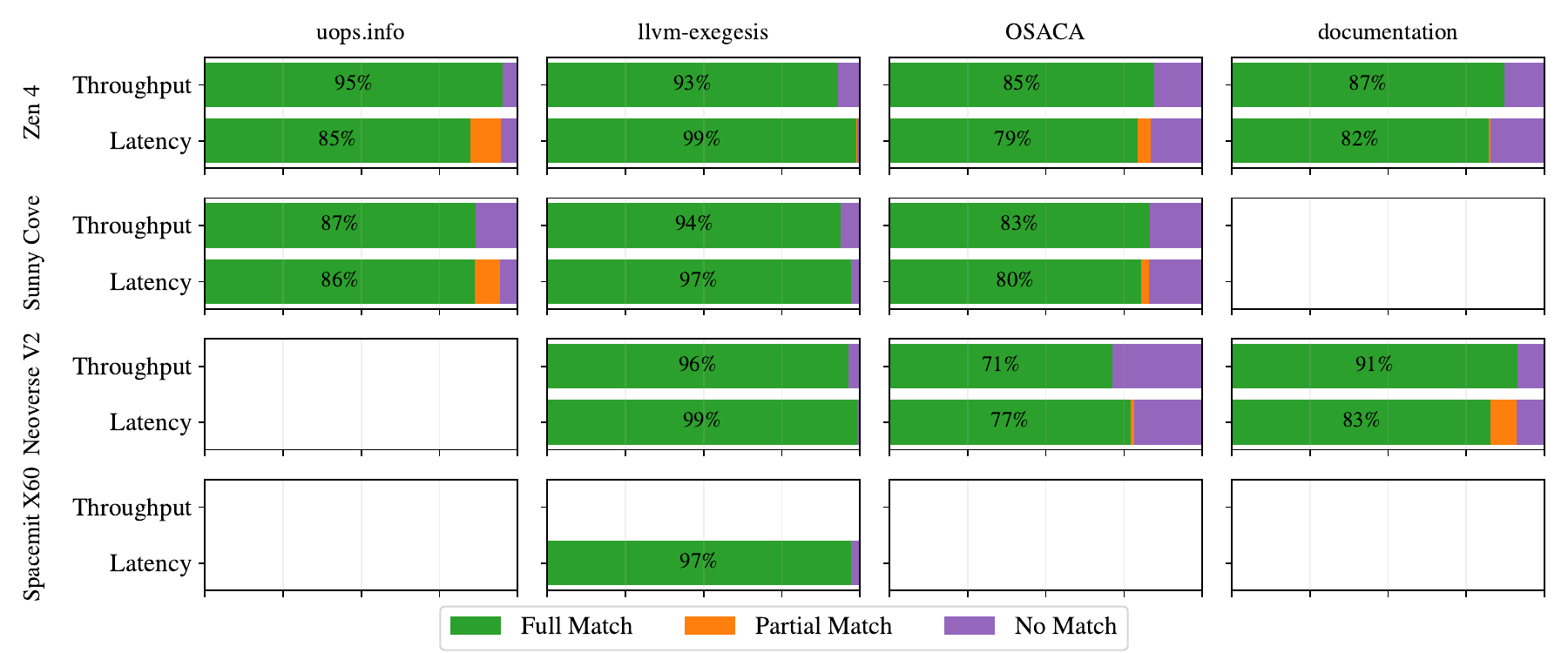}
    \caption{The matching of throughput and latency measurements for all used platforms in this work obtained by WINIC in comparison to reference data by other benchmark tools and vendor documentation. Empty cells indicate that the compared tool does not support this architecture or that no documentation was available.}
    \label{fig:res}
\end{figure*}



We compare the results obtained on our four test systems with a variety of other sources in Figure~\ref{fig:res}.
For each system, each source of reference values, and each metric, the figure contains a bar representing all instruction forms for which WINIC and the other source provide data.
Those instruction forms are categorized into \textit{full match, partial match}, and \textit{no match}.
If there are multiple matches for a measurement done by WINIC (e.g., because the reference covers a specific immediate value as operand), we combine the values of all matches into one.
Then we build a set of obtained metrics each for WINIC and the other source.
An instruction is counted as \textit{full match} if one of the sets is a subset of the other.
If the sets partially overlap, an instruction form is classified as \textit{partial match} while no overlap (i.e., two distinctive sets) results in a classification of \textit{no match}.
To account for small deviations, we use a tolerance of 10\% whenever comparing two values.
Table \ref{tab:res} in the appendix contains the absolute numbers on which this figure is based.

For x86 processors, uops.info provides a comprehensive database of latency and throughput values obtained using nanoBench.
For the Zen\,4 system, 95\% of the instruction forms are full or partial matches for both latency and throughput.
For the Sunny Cove system, 94\% and 87\% of the instruction form measurements match fully or partially in latency and throughput, respectively.

llvm-exegesis is the only microbenchmarking tool that supports all three architectures, though it crashed on the RISC-V system in throughput mode during our testing.
Overall, 93\% to 99\% of instruction forms match across all platforms.

The OSACA database was compiled manually from different sources including uops.info, documentation, and manual measurements using iBench and asmbench, therefore, the overall number of values is lower than of the other reference sources and it does not support RISC-V.
The proportion of non-matching values is higher compared to the other sources on all platforms.
Many of those non-matching values can be identified as register-to-register \texttt{MOV} instructions, which are often assumed to be not bound to any execution unit as the core can apply register renaming; however, in an isolated benchmark, WINIC measures a non-zero value for them.

Lastly, we compare WINIC's results to vendor documentation. 
Compared to the Arm Software Optimization Guide, 91\% of the instruction forms are full or partial matches for both latency and throughput.
For Zen\,4, AMD provides a latency and throughput overview in the Software Optimization Guide for the AMD Zen\,4 Microarchitecture~\cite{Zen4Opt}.
82\% of the instruction forms have matching latencies and 87\% have matching throughput values.
We noticed that vendor documentation does not necessarily contain data on all instruction forms.
The Zen\,4 sheet, for example, does not include the variants of arithmetic instructions with an immediate value as operand. 
For the other two microarchitectures, we did not find publicly available documentation of latency and throughput metrics.

\section{Case Studies}\label{sec:case_study}
To understand the value of considering operand-based latencies obtained by WINIC for in-core performance modeling, in the following, we investigate two different latency-bound loop kernels.
As OSACA does not support operand-based latency modeling at the time of writing, we manually adjust OSACA's output for the refined analyses and format them the same way as OSACA for the reader's convenience.

\subsection{Modified Schönauer DAXPY}
Fused-multiply-add operations as part of the limiting dependency chain throughout a streaming 1D-fashion as well as in multiple dimensions are common in a variety of scientific applications; among others, they can be found in stencil codes and linear algebra kernels, e.g., Fast Fourier Transform~(FFT)~\cite{FMAFFT}, matrix-matrix and matrix-vector multiplications (GEMM, DGEMM, SpGEMM, DMVM)~\cite{FMAMatrix}, and the BLAS algorithm AXPY~($y = ax + y$).
For simplicity, we look at a slightly modified Schönauer DAXPY kernel~(i.e., double-precision AXPY with three data arrays) that introduces a dependency between loop iterations (\texttt{a[i]} is dependent on \texttt{a[i-1]}), as shown in Listing~\ref{lst:schönauer}.

\begin{listing}[ht]
\begin{minted}
[
xleftmargin=20pt,
frame=lines,
framesep=2mm,
baselinestretch=1.2,
bgcolor=lighter-gray,
fontsize=\footnotesize,
linenos
]
{cpp}
double a[N], b[N], c[N];

for(int i=1; i<N; ++i) {
    a[i] =  a[i-1] + b[i] * c[i];
}
\end{minted}
\vspace{-2em}
\caption{A variant of the Schönauer DAXPY kernel with a read-after-write dependency.}
\label{lst:schönauer}
\end{listing}

The kernel was compiled for Neoverse-V2 using the Clang compiler at version 20.1.0 with flags \texttt{-mcpu=grace}, \texttt{-O3}, and \texttt{-fno-unroll-loops}. The resulting assembly code of the loop kernel can be found in Listing~\ref{lst:schönauer_asm}.

\begin{listing}[]
\begin{minted}
[
xleftmargin=20pt,
frame=lines,
framesep=2mm,
baselinestretch=1.2,
bgcolor=lighter-gray,
fontsize=\footnotesize,
linenos
]
{asm}
.LBB0_2:
    ldr     d1, [x9], #8
    ldr     d2, [x10], #8
    subs    x11, x11, #1
    fmadd   d0, d1, d2, d0
    str     d0, [x8], #8
    b.ne    .LBB0_2
\end{minted}
\vspace{-2em}
\caption{The resulting assembly code for the adjusted Schönauer DAXPY kernel in Listing~\ref{lst:schönauer} for the NVIDIA Grace chip.}
\label{lst:schönauer_asm}
\end{listing}

While the loading of the values \texttt{b[i]} and \texttt{c[i]} in lines 2--3, as well as the loop mechanics in lines 4 and 7 and the write-back into \texttt{a[i]} can overlap with all instructions of the other loop iterations, the FMA instruction \texttt{FMADD} shows an unresolvable dependency between the loops as register \texttt{d0} is read and written each iteration.
OSACA detects this dependency as expected and shows it in its output analysis,\footnote{Analysis also available at \url{https://godbolt.org/z/e3WWfb3eE}} as depicted in Figure~\ref{tab:osaca}.

\begin{figure*}[h]
    \centering
    \begin{adjustbox}{max width=\linewidth}
    \begin{tabular}{|r|*{16}{c|}c||c|c|l|}
    \hline
    line & 0 & 1 & 2 & 3 & 4 & 5 & 6 & 7 & 8 & 9 & 10 & 11 & 12 & 13 & 14 & 15 & 16 & CP & LCD & Instruction \\
    \hline
   1 &      &      &      &      &      &      &             &             &             &      &             &      &      &      &      &      &      &      &      &   \texttt{.LBB0\_2:} \\
   2 &      &      & 0.25 & 0.25 &      &      & 0.25        & 0.25        &             &      &             &      & 0.33 & 0.33 & 0.33 &      &      &  4.0 &      &   \texttt{ldr     d1, [x9], \#8} \\
   3 &      &      & 0.25 & 0.25 &      &      & 0.25        & 0.25        &             &      &             &      & 0.17 & 0.16 & 0.66 &      &      &      &      &   \texttt{ldr     d2, [x10], \#8} \\
   4 &      &      & 0.33 &      &      &      & 0.33        & 0.33        &             &      &             &      &      &      &      &      &      &      &      &   \texttt{subs    x11, x11, \#1} \\
   5 &      &      &      &      &      &      &             &             & 0.25        & 0.25 & 0.25        & 0.25 &      &      &      &      &      &  4.0 &  4.0 &   \texttt{fmadd   d0, d1, d2, d0} \\
   6 &      &      & 0.17 & 0.49 &      &      & 0.17        & 0.17        &             &      &             &      & 0.50 & 0.50 &      & 0.50 & 0.50 &  0.0 &      &   \texttt{str     d0, [x8], \#8} \\
   7 & 0.50 & 0.50 &      &      &      &      &             &             &             &      &             &      &      &      &      &      &      &      &      &   \texttt{b.ne    .LBB0\_2} \\
    \hline
     sum & 0.50 & 0.50 & 1.00 & 0.99 &  &  & 1.00 & 1.00 & 0.25 & 0.25 & 0.25 & 0.25 & 1.00 & 1.00 & 1.00 & 0.50 & 0.50 & 8.0 & \textcolor{red}{4.0} & \\
    \hline
    \end{tabular}
    \end{adjustbox}
    \caption{OSACA analysis for the adjusted Schönauer DAXPY kernel in Listing~\ref{lst:schönauer_asm}: The limiting bottleneck is highlighted in red and is expected to be $4$\,cycles/iteration due the loop-carried dependency on the \texttt{FMADD} instruction.}
    \label{tab:osaca}
\end{figure*}

\begin{figure*}[ht]
    \centering
    \begin{adjustbox}{max width=\linewidth}
    \begin{tabular}{|r|*{16}{c|}c||c|c|l|}
    \hline
    line & 0 & 1 & 2 & 3 & 4 & 5 & 6 & 7 & 8 & 9 & 10 & 11 & 12 & 13 & 14 & 15 & 16 & CP & LCD & Instruction \\
    \hline
   1 &      &      &      &      &      &      &             &             &             &      &             &      &      &      &      &      &      &      &      &   \texttt{.LBB0\_2:} \\
   2 &      &      & 0.25 & 0.25 &      &      & 0.25        & 0.25        &             &      &             &      & 0.33 & 0.33 & 0.33 &      &      &  4.0 &      &   \texttt{ldr     d1, [x9], \#8} \\
   3 &      &      & 0.25 & 0.25 &      &      & 0.25        & 0.25        &             &      &             &      & 0.17 & 0.16 & 0.66 &      &      &      &      &   \texttt{ldr     d2, [x10], \#8} \\
   4 &      &      & 0.33 &      &      &      & 0.33        & 0.33        &             &      &             &      &      &      &      &      &      &      &      &   \texttt{subs    x11, x11, \#1} \\
   5 &      &      &      &      &      &      &             &             & 0.25        & 0.25 & 0.25        & 0.25 &      &      &      &      &      &  4.0 &  2.0 &   \texttt{fmadd   d0, d1, d2, d0} \\
   6 &      &      & 0.17 & 0.49 &      &      & 0.17        & 0.17        &             &      &             &      & 0.50 & 0.50 &      & 0.50 & 0.50 &  0.0 &      &   \texttt{str     d0, [x8], \#8} \\
   7 & 0.50 & 0.50 &      &      &      &      &             &             &             &      &             &      &      &      &      &      &      &      &      &   \texttt{b.ne    .LBB0\_2} \\
    \hline
     sum & 0.50 & 0.50 & 1.00 & 0.99 &  &  & 1.00 & 1.00 & 0.25 & 0.25 & 0.25 & 0.25 & 1.00 & 1.00 & 1.00 & 0.50 & 0.50 & 8.0 & \textcolor{red}{2.0} & \\
    \hline
    \end{tabular}
    \end{adjustbox}
    \caption{Refined performance analysis for the kernel in Listing~\ref{lst:schönauer_asm}: Marked in red, the estimated runtime is now matching the measurements with $2$ cycles/iteration due to the new operand-based latency values.}
    \label{tab:osaca_corr}
\end{figure*}

Although the maximum throughput would be $1$\,cycle, as that is the highest port pressure on ports 2, 3, 6, 7, 12, 13 and 14, the LCD report correctly reveals that the loop is expected to be bound by the loop-carried dependency of the \texttt{FMADD} instruction.
OSACA (as well as llvm-mca) reports a latency of $4$\,cycles for the \iform{FMADD\,(FPR64,\,FPR64,\,FPR64)} instruction form, i.e., a minimum of $4$\,cycles per loop iteration is needed. However, when benchmarking this kernel on a NVIDIA Grace Superchip, we measure only $2$\,cycles per iteration, i.e., the initial model by OSACA is off by a factor of $2$.
To investigate why, we use WINIC to measure the operand-based latencies of the \texttt{FMADD} as follows:

\begin{listing}[H]
\begin{minted}
[
xleftmargin=20pt,
frame=lines,
framesep=2mm,
baselinestretch=1.2,
bgcolor=lighter-gray,
fontsize=\footnotesize,
linenos
]
{text}
winic LAT -f 3.2 -i FMADDDrrr
 FMADDDrrr(1<FPR64> -> 0<FPR64>)  [4.01;4.01]
 FMADDDrrr(2<FPR64> -> 0<FPR64>)  [4.01;4.01]
 FMADDDrrr(3<FPR64> -> 0<FPR64>)  [2.01;2.01]
 FMADDDrrr(Reg<FPCR> -> 0<FPR64>) [ERR_NO_HLPR]
\end{minted}
\vspace{-2em}
\caption{WINIC output for measuring the latency of \iform{FMADD\,(FPR64,\,FPR64,\,FPR64)} on a NVIDIA Grace Superchip.}
\label{lst:fmadd_out}
\end{listing}

The output shows that \texttt{FMADD} has four operand-based latencies. 
Each line in the output describes one dependency indicated by an arrow and the minimum and maximum cycles measured by WINIC in square brackets.
While the dependencies from operands 1 and 2---holding the multiplicand and multiplier---to operand 0, i.e., the destination register, take $4$\,cycles, the dependency from the addend (operand~3) to the destination register only takes $2$\,cycles.
Measuring the last dependency from the floating point control register~(FPCR) to operand~0 fails, as it is of asymmetric type and, therefore, would need an additional instruction form\footnote{WINIC can, in fact, not measure this specific operand-based latency as no instruction form of the AArch64 instruction set has an operand-based latency of type \ltype{FPR64}{FPCR}. This is because the FPCR controls floating point rounding modes, overflow behavior, etc., and can only be set by the \iform{MSR\,(FPCR,\,r64)} (Move to Special Register) instruction form. 
This being said, for the missing value to affect performance, one would need a loop kernel where the floating point behavior is altered every iteration, which is highly unlikely. Therefore, this value can be considered irrelevant.}
as described in Section~\ref{ssec:measuring_latency}. 
The difference in the operand-based latency measurements reveals the reason for the mismatch between the OSACA prediction and the empirical result.
We can update OSACA's performance model with the new insight gained by WINIC and introduce operand-based latencies for the in-core performance analysis.
Figure~\ref{tab:osaca_corr} shows the refined performance model for the same kernel from Listing~\ref{lst:schönauer_asm}.

As the loop-carried dependency in the \texttt{FMADD} instruction is from operand~3 to operand~0, the latency gets updated to 2\,cycles and the overall prediction of the kernel becomes 2\,cycles per iteration, matching the measurement and refining the initial model. 

\subsection{Modified 2D-5pt Gauss-Seidel}\label{sec:gauss-seidel}
The 2D-5pt Gauss-Seidel algorithm is an iterative solver for linear partial differential equations~(PDEs), such as the Poisson or Laplace equation, and can be used as multi-grid smoother~\cite{barrett1994templates}.
In this case study we will look at a slightly modified Gauss-Seidel method with an individual weight for each cell rather than an equal scaling factor.
Thus, the weights must be kept in a separate array.
The code can be seen in Listing~\ref{lst:gauss-seidel}:

\begin{listing}[h]
\begin{minted}
[
xleftmargin=20pt,
frame=lines,
framesep=2mm,
baselinestretch=1.2,
bgcolor=lighter-gray,
fontsize=\footnotesize,
linenos
]
{cpp}
double a[NI][NK], b[NI][NK];

for (int i=1; i<NI-1; ++i) {
  for (int k=1; k<NK-1; ++k) {
    a[i][k] = (
                    a[i+1][k]
      + a[i][k-1] +           + a[i][k+1]
                  + a[i-1][k]
    ) / b[i][k];
  }
}
\end{minted}
\vspace{-2em}
\caption{The 2D-5pt Gauss-Seidel method using individual weights for each cell.}
\label{lst:gauss-seidel}
\end{listing}

We compile the code using armclang 20.1.0 with flags \texttt{-O3 -mcpu=grace -funroll-loops -fno-builtin} and analyze the emitted assembly using OSACA as shown in Figure~\ref{fig:gs-osaca-1}. 
We can see that---despite the maximum possible reciprocal throughput of $5$\,cycles per iteration---the code is strongly bound by the loop-carried dependency, specifically the latency of the divide instruction, making up for $24$ out of the $36$\,cycles for the two-time unrolled loop kernel, as shown in column ``LCD''.

 \begin{figure*}[tp]
     \centering
     \begin{adjustbox}{max width=\linewidth}
     \begin{tabular}{|r|*{18}{c|}c||c|c|l|}
     \hline
     line & 0 & 1 & 2 & 3 & 4 & 5 & 6 & 7 & 8 & 8D & 9 & 9D & 10 & 11 & 12 & 13 & 14 & 15 & 16 & CP & LCD & Instruction \\
     \hline
    1 &      &      &      &      &      &      &             &             &      &       &      &      &       &      &      &      &      &      &      &      &      & \texttt{.LBB6\_42:} \\
    2 &      &      & 0.17 & 0.167& 0.17 & 0.17 & 0.17        & 0.17        &      &       &      &      &       &      &      &      &      &      &      & 1.0  &      & \texttt{add x18, x27, x17}  \\
    3 &      &      & 0.17 & 0.167& 0.17 & 0.17 & 0.17        & 0.17        &      &       &      &      &       &      & 0.17 & 0.16 & 0.66 &      &      &      &      & \texttt{add x0, x24, x17}   \\
    4 &      &      & 0.08 & 0.097& 0.33 & 0.33 & 0.09        & 0.09        &      &       &      &      &       &      &      &      &      &      &      &      &      & \texttt{add x1, x22, x17}   \\
    5 &      &      & 0.25 & 0.26 &      &      & 0.24        & 0.25        &      &       &      &      &       &      &      &      &      &      &      &      &      & \texttt{sub x16, x16, \#2}  \\
    6 &      &      &      &      &      &      &             &             &      &       &      &      &       &      & 0.33 & 0.33 & 0.33 &      &      &  6.0 &      & \texttt{ldp d1, d2, [x18]}  \\
    7 &      &      & 0.00 & 0.00 & 0.50 & 0.50 & 0.00        & 0.00        &      &       &      &      &       &      &      &      &      &      &      &      &      & \texttt{add x18, x21, x17}  \\
    8 &      &      & 0.17 & 0.49 &      &      & 0.17        & 0.17        &      &       &      &      &       &      &      &      &      &      &      &      &      & \texttt{add x17, x17, \#16}  \\
    9 &      &      & 0.33 &      &      &      & 0.33        & 0.33        &      &       &      &      &       &      &      &      &      &      &      &      &      & \texttt{cmp x20, x16}   \\
    10&      &      &      &      &      &      &             &             & 0.25 &       & 0.25 & 0.25 &       & 0.25 &      &      &      &      &      &  2.0 &  2.0 & \texttt{fadd d0, d0, d1}    \\
    11&      &      &      &      &      &      &             &             &      &       &      &      &       &      & 0.33 & 0.33 & 0.33 &      &      &      &      & \texttt{ldur d1, [x18, \#-8]}   \\
    12&      &      &      &      &      &      &             &             & 0.25 &       & 0.25 & 0.25 &       & 0.25 &      &      &      &      &      &  2.0 &  2.0 & \texttt{fadd d0, d0, d1}    \\
    13&      &      &      &      &      &      &             &             &      &       &      &      &       &      & 0.33 & 0.33 & 0.33 &      &      &      &      & \texttt{ldr d1, [x0, \#8]}   \\
    14&      &      &      &      &      &      &             &             & 0.25 &       & 0.25 & 0.25 &       & 0.25 &      &      &      &      &      &  2.0 &  2.0 & \texttt{fadd d0, d0, d1}    \\
    15&      &      &      &      &      &      &             &             &      &       &      &      &       &      & 0.33 & 0.33 & 0.33 &      &      &      &      & \texttt{ldp d1, d3, [x1]}   \\
    16&      &      &      &      &      &      &             &             & 0.50 &  2.50 &      & 0.50 &  2.50 &      &      &      &      &      &      & 12.0 & 12.0 & \texttt{fdiv d0, d0, d1}    \\
    17&      &      &      &      &      &      &             &             &      &       &      &      &       &      & 0.33 & 0.33 & 0.33 &      &      &      &      & \texttt{ldr d1, [x18]}  \\
    18&      &      &      &      &      &      &             &             &      &       &      &      &       &      & 0.50 & 0.50 &      & 0.50 & 0.50 &      &      & \texttt{stur d0, [x18, \#-16]}  \\
    19&      &      &      &      &      &      &             &             & 0.25 &       & 0.25 & 0.25 &       & 0.25 &      &      &      &      &      &  2.0 &  2.0 & \texttt{fadd d0, d0, d2}    \\
    20&      &      &      &      &      &      &             &             & 0.00 &       & 0.50 & 0.00 &       & 0.50 &      &      &      &      &      &  2.0 &  2.0 & \texttt{fadd d0, d0, d1}    \\
    21&      &      &      &      &      &      &             &             &      &       &      &      &       &      & 0.00 & 0.00 & 1.00 &      &      &      &      & \texttt{ldr d1, [x0, \#16]} \\
    22&      &      &      &      &      &      &             &             & 0.00 &       &      & 0.00 &       & 0.50 &      &      &      &      &      &  2.0 &  2.0 & \texttt{fadd d0, d0, d1}    \\
    23&      &      &      &      &      &      &             &             & 0.50 &  2.50 &      & 0.50 &  2.50 &      &      &      &      &      &      & 12.0 & 12.0 & \texttt{fdiv d0, d0, d3}    \\
    24&      &      &      &      &      &      &             &             &      &       &      &      &       &      & 0.50 & 0.50 &      & 0.50 & 0.50 &  0.0 &      & \texttt{stur d0, [x18, \#-8]}   \\
    25& 0.50 & 0.50 &      &      &      &      &             &             &      &       &      &      &       &      &      &      &      &      &      &      &      & \texttt{b.ne .LBB6\_42} \\
     \hline
      sum & 0.50 & 0.50 & 1.16 & 1.171 & 1.16 & 1.16 & 1.16 & 1.17 & 2.00 & 5.00 & 2.00 & 2.00 & 5.00 & 2.00 & 2.67 & 2.67 & 2.67 & 1.00 & 1.00 & 43.0 & \textcolor{red}{36.0} & \\
     \hline
     \end{tabular}
     \end{adjustbox}
     \caption{OSACA analysis for the modified 2D-5pt Gauss-Seidel kernel in Listing~\ref{lst:gauss-seidel}. The expected bottleneck limiting the overall runtime is marked in red.}
     \label{fig:gs-osaca-1}
 \end{figure*}

However, when benchmarking this Gauss-Seidel kernel and measuring the cycles using hardware performance counters, we observe a runtime of $28.28\,\textrm{cycles}$ per two assembly iterations, significantly deviating from OSACA's prediction.
WINIC reveals that the \iform{FDIV\,(FPR64,\,FPR64,\,FPR64)} instruction form in fact has two different operand-based latencies:

\begin{listing}[H]
\begin{minted}
[
xleftmargin=20pt,
frame=lines,
framesep=2mm,
baselinestretch=1.2,
bgcolor=lighter-gray,
fontsize=\footnotesize,
linenos
]
{text}
winic LAT -f 3.2 -i FDIVDrr
 FDIVDrr(1<FPR64> -> 0<FPR64>)  [7.01;7.01]
 FDIVDrr(2<FPR64> -> 0<FPR64>)  [14.5;14.5]
\end{minted}
\vspace{-2em}
\caption{WINIC output for measuring the latency of \iform{FDIV\,(FPR64,\,FPR64,\,FPR64)} on a NVIDIA Grace Superchip.}
\label{lst:fdiv_out}
\end{listing}

Furthermore, the Arm Software Optimization Guide~\cite{NeoverseSOG} reveals that the effective latency of the \texttt{FDIV} instruction is increased by one cycle as the core cannot apply region-based fast forwarding to other pipelines.
As the dependency chain in the Gauss-Seidel kernel spans from the third operand to the first operand (\texttt{d0} in line 14 for the \texttt{FDIV} in line 16 and \texttt{d0} in line 22 for the \texttt{FDIV} in line 23 in Figure~\ref{fig:gs-osaca-1}, respectively), this insight requires us to change the latency for the instruction form to $7+1=8\,\textrm{cy}$ in this example.
The latencies for the refined loop-carried dependency chain now sum up to $2+2+2+8+2+2+2+8=28$ (lines 10, 12, 14, 16, 19, 20, 22, and 23), matching the observed measurement of $28.28\,\textrm{cy}$ exactly (deviation of 1\%).

\section{Conclusion}\label{sec:conclusion}
\subsection{Summary}
WINIC is a platform-independent, cycle-accurate microbenchmarking tool running in user-space for performance analysis and modeling.
It can automatically benchmark throughput and latency of most instruction forms available in LLVM on x86, AArch64, and RISC-V.
Furthermore, WINIC can break dependencies for more accurate throughput benchmarks if needed, and it is capable of measuring operand-based latencies in instruction forms, which, to the best of our knowledge, no other tool prior was able to for AArch64 and RISC-V instruction sets.
To prove WINIC's accuracy, we conducted an extensive comparison with existing data from state-of-the-art microbenchmarking tools and available documentation, revealing full or partial alignment for 71\%--99\% of all measured instructions.
This shows that WINIC is a versatile multi-platform benchmarking tool that is in many cases more accurate than existing vendor documentation.

Finally, we showed that WINIC's ability to measure operand-based latencies can be used to build and refine existing in-core performance models.
This was demonstrated using two scientific kernels on the NVIDIA Grace CPU Superchip, in which the refined models significantly improve the accuracy of the predictions.

\subsection{Future Work}\label{ssec:future_work}
In the near future, we plan to apply evolutionary algorithms to derive the port usage of instructions based on the measurements.
While WINIC currently only constructs latency chains between up to two types of instructions and, therefore, can only predict a latency range for some operand-based latencies, larger combinations of instruction types might yield exact results for all instruction forms involved.
Furthermore, future work will also consider techniques to accurately measure region-based forwarding penalties on Arm-based platforms.

\section*{Acknowledgment}
This work was partly funded by BMFTR through the DAREXA-F project.

\bibliographystyle{IEEEtran}
\bibliography{bibtex}

\clearpage
\onecolumn
\appendix
\newcommand{\m}{ 
    }
    \newcommand{\ms}{ 
    }

\setcounter{table}{0}
\renewcommand{\thetable}{A\arabic{table}}

\section{Comparison to Existing Data}\label{ssec:appendix}
Table \ref{tab:res} shows the absolute match numbers from the comparison in Figure \ref{fig:res}.
The last column shows the total number of instruction forms for which WINIC was able to obtain a throughput or latency value, respectively.

\begin{table*}[h]
    \centering
    \caption{The number of obtained throughput and latency measurements for all used platforms in this work and the matching of these measurements to reference data by other benchmark tools and vendor documentation. Empty cells indicate that the compared tool does not support this architecture or that no documentation was available.}
    \label{tab:res}
    \begin{adjustbox}{max width=\textwidth}
    \begin{tabular}{c|c||cc|cc|cc|cc||cc}
        && \multicolumn{2}{c|}{uops.info} & \multicolumn{2}{c|}{llvm-exegesis} & \multicolumn{2}{c|}{OSACA} & \multicolumn{2}{c||}{documentation} & \multicolumn{2}{N}{obtained instr.} \\
        && TP & LAT & TP & LAT & TP & LAT & TP & LAT & TP & LAT \\
        \midrule
        & full match                              & 12765 & 11205 & 12484 & 5528 & 1418 & 1305 & 9712 & 9148 & & \\
        & partial match                           & 0 & 1318 & 0 & 3 & 0 & 69 & 0 & 50 & 13605 & 13689 \\
        \multirow{-3}{*}{Zen\,4} & no match       & 618 & 684 & 940 & 52 & 256 & 272 & 1426 & 1927 & & \\
        \hline
        \rowcolor{lighter-gray}
        & full match                              & 11504 & 11282 & 12507 & 5402 & 1942 & 1810 &  &  & & \\
        \rowcolor{lighter-gray}
        & partial match                           & 13 & 1045 & 0 & 0 & 0 & 60 &  &  & 13489 & 13528 \\
        \rowcolor{lighter-gray}
        \multirow{-3}{*}{Sunny Cove} & no match   & 1764 & 724 & 803 & 150 & 390 & 382 & \m &  & & \\
        \hline
        & full match                              &  &  & 3304 & 3133 & 600 & 658 & 3320 & 2950 & & \\
        & partial match                           &  &  & 0 & 0 & 0 & 8 & 0 & 289 & 3715 & 3612 \\
        \multirow{-3}{*}{Neoverse V2} & no match  & \m &  & 123 & 31 & 242 & 186 & 311 & 322 & &  \\
        \hline
        \rowcolor{lighter-gray}
        & full match                              &  &  &  & 102 &  &  &  &  & & \\
        \rowcolor{lighter-gray}
        & partial match                           &  &  &  & 0 &  &  &  &  & 707 & 712 \\
        \rowcolor{lighter-gray}
        \multirow{-3}{*}{Spacemit X60} & no match  & \m &  &  & 3 & \m &  & \m &  & & \\
    \end{tabular}
    \end{adjustbox}
\end{table*}

\end{document}

\typeout{get arXiv to do 4 passes: Label(s) may have changed. Rerun}